# Bayesian Characterization of Uncertainties Surrounding Fluvial Flood Hazard Estimates

Sanjib Sharma[1,*], Ganesh Raj Ghimire[2], Rocky Talchabhadel[3], Jeeban Panthi[4], Benjamin Seiyon Lee[5], Fengyun Sun[6], Rupesh Baniya[7], Tirtha Raj Adhikari[8]

[1]Earth and Environmental Systems Institute, The Pennsylvania State University, University Park, PA, USA
[2]IIHR Hydroscience and Engineering, The University of Iowa, Iowa City, Iowa, USA
[3]Texas A&M AgriLife Research, Texas A&M University, El Paso, TX, USA
[4]Department of Geosciences, University of Rhode Island, Kingston, RI, USA
[5]Department of Statistics, The George Mason University, Fairfax, VA, USA
[6]School of Ecological and Environmental Sciences, East China Normal University, Shanghai, 200241, China
[7]Institute of Engineering, Tribhuvan University, Lalitpur, 44600, Nepal
[8]Central Department of Hydrology and Meteorology, Tribhuvan University, Kirtipur, 44618, Nepal
*Email: sanjibsharma66@gmail.com

## Abstract

Fluvial floods drive severe risk to riverine communities. There is a strong evidence of increasing flood hazards in many regions around the world. The choice of methods and assumptions used in flood hazard estimates can impact the design of risk management strategies. In this study, we characterize the expected flood hazards conditioned on the uncertain model structures, model parameters and prior distributions of the parameters. We construct a Bayesian framework for river stage return level estimation using a nonstationary statistical model that relies exclusively on Indian Ocean Dipole Index. We show that ignoring uncertainties can lead to biased estimation of expected flood hazards. We find that the considered model parametric uncertainty is more influential than model structures and model priors. Our results highlight the importance of incorporating uncertainty in river stage estimates, and are of practical use for informing water infrastructure designs in a changing climate.

# 1. Introduction

Floods are among the most devastating climate-related disasters across the globe (Cook et al., 2018, Winsemius et al. 2016, Hirabayashi et al. 2013). Flooding poses substantial risks to life and property destroying civil infrastructures, causing embankment overtopping and breaching, and displacing settlements in downstream lowland areas. The performance of strategies to manage these risks relies on sound understanding of extreme flood characteristics in a changing climate.

Statistical distributions of extreme floods have been widely used to inform the design and operation of water infrastructures such as bridges, dams and reservoirs (Obeysekera and Salas 2014). For instances, bridge engineers rely on extreme flood records to determine bridge freeboard requirements; hydroelectric dam operators rely on extreme water level estimates to manage the power supply and reduce the frequency of spillage; drainage engineers rely on extreme rainfall characteristics to size stormwater infrastructure systems; and floodplain managers utilize flood peak records to delineate floodplain boundaries that are most vulnerable to floods. The design standard for many water resources practices assume that the statistical distribution of flood records is stationary, i.e., the occurrence of extreme flood event is not expected to change significantly over time (Jakob 2013, Milly et al. 2008, Obeysekera and Salas 2014, Mallakpour et al. 2019, Pralle 2019).

There is considerable physical evidence of increasing both the frequency and intensity of extreme events in several parts of the world due to global atmospheric warming and/or local anthropogenic impacts, such as land-use land-cover changes and reservoir regulations (Yang et al. 2013, Francios et al. 2019, Winsemius et al. 2016). However, traditional infrastructure design practice neglects the potential changes in the hydrological conditions of flood records, hereafter called as nonstationarity. Given the nonstationarity in flood records, a stationary assumption may lead to unreliable estimation of design floods (Sarhadi and Soulis 2017, Tan and Gan 2015, Salas et al. 2018, Dong et al. 2019). Thus, if the properties of flood events have changed over time, current engineering standards may yield poor infrastructure design choices.

Several previous studies (Cheng and AghaKouchak 2015, Sarhadi and Soulis 2017; Ouarda and Charron 2019) have provided important new insights on the impacts of climate change on hydroclimatic extremes, but are silent on the characterization of influential uncertainty sources on extreme flood estimates. Extreme flood estimates are inherently uncertain due to several reasons (Neppel et al. 2010, Meresa and Romanowicz 2017, Parkes and Demeritt 2016, Qi et al. 2016,

Steinschneider et al. 2012, Hu et al. 2019, Vidrio-Sahagun et al. 2020). First, extreme floods are characterized by limited data records and require modeling where the hydrologic stations are most sparse (Wong et al. 2018). Second, there is limited additional information supplementary to the streamflow time series to be integrated as prior information on the parameters of extreme value distributions (Lee et al. 2017). Third, there is no strong consensus among experts and/or decision makers regarding the choice of model structure to use as a best practice to inform infrastructure design as well as for climate change impact assessments (Wong et al. 2018). Most of the previous efforts on uncertainty characterization have been focused on analyzing the individual and/or combined effect of few uncertainty sources on extreme flood estimates (Merz and Thieken 2005, Liang et al. 2011, Hue et al. 2019, Debele et al. 2017, Vidrio-Sahagún 2020). However, studies are limited to understand how uncertainty from different sources interacts and propagates in extreme value analysis, and hence the relation between individual uncertainty and the total uncertainty is unclear. Comprehensive uncertainty characterization in extreme flood estimates can be critical to provide decision-relevant extreme information for which the infrastructure will be designed to withstand during their lifetime.

This study presents a i) formal statistical framework to incorporate nonstationarity into extreme river stage estimates and ii) use this framework to identify the most influential uncertainty sources and their interactions in river stage return level estimates. The proposed framework is based on the GEV distribution combined with the Bayesian-based Markov chain approach for predictive intervals. We perform sensitivity analysis of extreme river stage on specification of different uncertainty sources, including model priors, model structures and prior distributions of the parameters. Overall, this study highlights the key avenues to advance our understanding of the climate-induced changes in the hydroclimatic records, improve the estimation of extreme river stage and discharge, and present the methodology that can potentially be integrated into design concept or hazard management schemes to withstand the specific standard of protection.

## 2. Materials and Methods

### 2.1. Study area

We choose major river basins in the central Himalayan region Nepal, where frequent and severe flooding are the major concern (Kattelmann 2003, Mool et al. 2001). The selected hydrologic stations are at Arughat, Asaraghat, Benighat, Rabuwabazar, Bagasoti, Betrawati, Chisapani, Narayanghat and BangaBelgaon (Fig. 1, Table 1). These stations are monitored by the Department of Hydrology and Meteorology (DHM), Nepal. We select these stations on the basis that they have longest publicly available data record, no period of missing data, and are categorized as "good data quality" by the DHM. We select central Himalayan region for several reasons. The Himalayas are referred to as the water towers of Asia and provide water resources to millions of people (Sharma et al., 2019). Extreme river gauge records from the selected hydrologic stations have been widely used to inform the design of critical infrastructures (Shrestha et al. 2019, Sharma et al. 2019). For instance, the river gauge records from Arughat station are used to inform the design of the Budhi Gandaki hydroelectric project, which is the largest proposed hydroelectric infrastructure in Nepal with a dam height of 263 m and a capacity of 1200 Megawatts (Gyawali 2019). In addition, a network of operational and proposed hydroelectric infrastructure across the country (Fig. 1) clearly depicts the need of comprehensive assessment of river flow characteristics to inform infrastructure design decisions in a changing climate (Farinotti et al. 2019). However, most of the previous studies on hydrologic extremes of central Himalayas have relied on short streamflow records (Gautam and Acharya 2012), focused only on particular region (Devkota and Gyawali 2015, Shrestha et al. 2014), and/or are based on stationarity assumption in river flow records (Mishra et al. 2009, Dhital and Kayastha 2013). Particularly in the context of central Himalayan Region Nepal, this study for the first time (to the authors knowledge) explores the temporal changes in flood peak distributions based on the longest publicly available water stage and discharge records in major river basins.

**Table 1**: Characteristics of the selected hydrologic stations in central Himalayas. These stations are monitored by the Department of Hydrology and Meteorology, Nepal.

| Station ID | Station Name | River Basin | Lat | Long | Basin Area (km$^2$) | Station Record (year) | River Gage Record (m) Max | Min | Mean |
|---|---|---|---|---|---|---|---|---|---|
| 260 | BangaBelgaon | Seti | 28.97 | 81.14 | 7366 | 1975-2013 | 10 | 5.3 | 7.36 |
| 240 | Asaraghat | Karnali | 28.95 | 81.44 | 21438 | 1963-2013 | 8.64 | 3.53 | 5.20 |
| 250 | Benighat | Karnali | 28.96 | 81.11 | 23229 | 1963-2010 | 11.8 | 3.76 | 8.08 |
| 280 | Chisapani | Karnali | 28.64 | 81.29 | 45857 | 1963-2015 | 15.2 | 6.96 | 10.36 |
| 350 | Bagasotigaon | West Rapti | 27.85 | 82.79 | 3551 | 1976-2015 | 9.84 | 3.49 | 5.52 |
| 450 | Narayanghat | Narayani | 27.69 | 84.43 | 32099 | 1977-2015 | 9.4 | 4.0 | 7.67 |
| 445 | Arughat | Budhi Gandaki | 28.04 | 84.81 | 3960 | 1969-2015 | 7.2 | 3.6 | 4.55 |
| 447 | Betrawati | Trishuli | 27.96 | 85.18 | 4110 | 1977-2015 | 5.66 | 2.95 | 4.07 |
| 670 | Rabuwabazar | Dudh Kosi | 27.27 | 86.66 | 3650 | 1973-2014 | 10 | 2.57 | 5.55 |

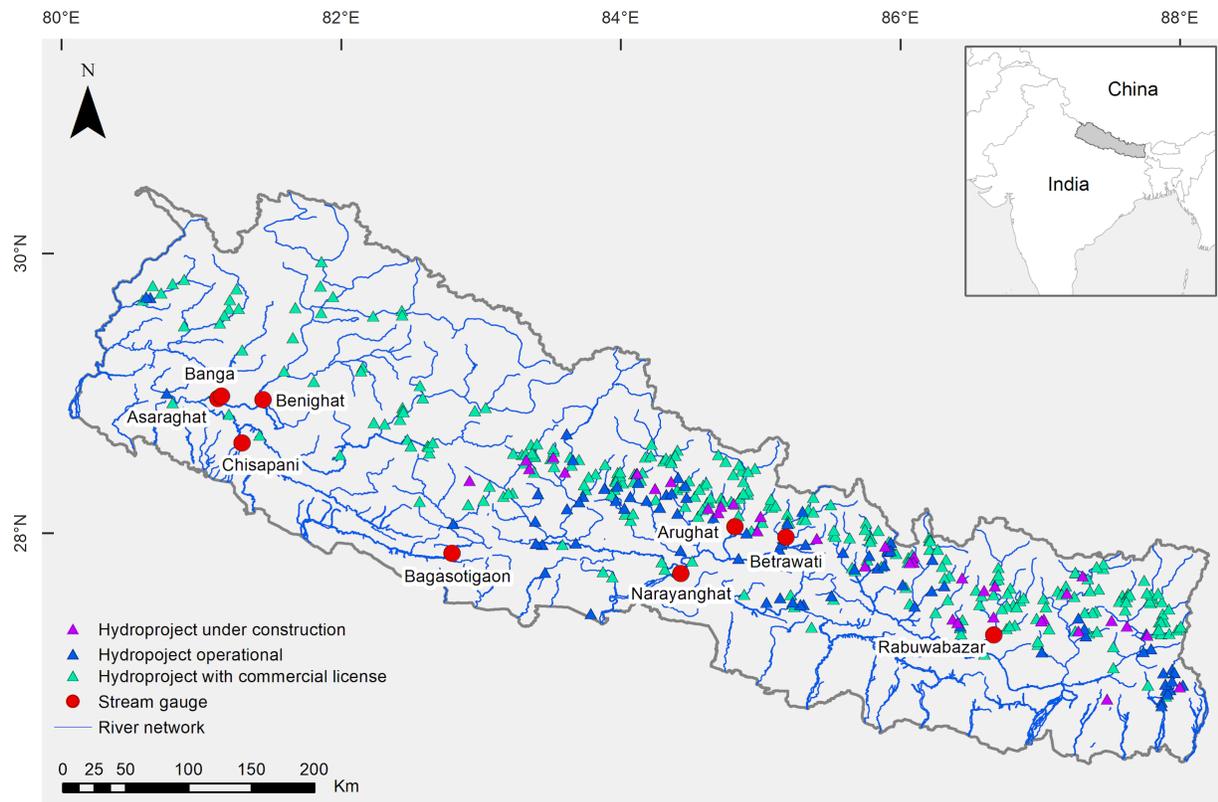

**Figure 1**. Map of central Himalayan country, Nepal. The map shows the stream network, location of the selected hydrologic stations, and location of hydropower stations (operational, under construction, and with commercial license). Major land cover types in Nepal are forest (39.1%), agriculture (29.8%), barren (10.7%), and snow/glaciers (8.2%) (Uddin et al. 2015). There are currently eighty operational hydroelectric projects (blue triangles) while two hundred and seventy-four other hydro projects (light green triangles) have received licenses for power generation (Alam et al. 2017). Detailed information on hydroelectric projects can be obtained from Department of Electricity Development, Nepal (https://www.doed.gov.np/ ).

## 2.2. River stage statistical modeling

Both the gradually varying long-term trends and abrupt changes can cause nonstationarity in annual river stage records. The validity of temporal stationarity assumption in observed gauge records is examined interms of abrupt changes in mean, and monotonic trends. We use the non-parametric rank-based Pettit test to detect change-points in river stage records (Pettitt 1979). For the gauge records that do not present a statistically significant change point in mean, nonparametric Mann-Kendall test (Kendall 1975) is used to examine the presence of monotonically increasing/decreasing trend in the historical records. If a change-point in mean is detected, annual

Under review in Hydrological Sciences Journal

maxima river stage timeseries is split into two subseries and the trend analysis is performed on each of the two subseries (Villarini and Smith 2010, Villarini et al. 2011). In this study, we assume there is a single, major change-point to avoid dividing the limited available instantaneous river gauge records (Table 1) into several subseries. Bayesian models capable of detecting multiple change-points could be investigated in future for more meaningful trend analysis (Ruggieri 2013).

We model the extreme river stage with a nonstationary GEV distribution and is recommended by several previous studies in the central Himalayan region (Mishra et al. 2009, Bohlinger and Sorteberf 2018, Kindermann et al. 2020). The GEV has the advantage that the potential nonstationarity in annual maxima river gauge records can be included explicitly by using covariate-dependent parameters. Thus, the expected return river stage under nonstationary assumption provides the return-probability of being equaled or exceeded in a particular year. However, under stationary assumption, expected return river stage has the same probability of being equaled or exceeded in any given year. The GEV distribution has the location parameter ($\mu$), the scale parameter ($\sigma$), and the shape parameter ($\xi$) to specify the center, spread, and tail heaviness, respectively. The probability density function of the GEV distribution is:

$$f(x) = \begin{cases} \frac{1}{\sigma}\left(1 + \xi\frac{w-\mu}{\sigma}\right)^{-1-\frac{1}{\xi}} e^{-\left(1+\xi\frac{w-\mu}{\sigma}\right)^{-\frac{1}{\xi}}}, & for\ 1 + \xi\frac{w-\mu}{\sigma} > 0 \\ \frac{1}{\sigma} e^{\frac{\mu-w}{\sigma} - e^{\frac{\mu-w}{\sigma}}}, & for\ \xi = 0 \\ 0, & otherwise \end{cases} \quad (1)$$

Based on the shape parameter, the GEV can take one of three forms: Gumbel, or light tailed, when $\xi$ is zero; Fréchet, or heavy tailed, if $\xi$ is positive; and Weibull, or bounded, when $\xi$ is negative.

Availability of limited observational records can constrain extreme floods, leading to a reliance on a simpler extreme value model with fewer parameters. We incorporate potential nonstationary into the GEV model by allowing the location parameter ($\mu$) to be a function of covariate (Cheng and AghaKouchak 2015; De Paola et al. 2018):

$$\mu(t) = \mu_0 + \mu_1 \varphi(t), \quad (2)$$

where $\mu_0$ is the unknown constant regression parameter, and $\varphi(t)$ is the time-series covariate that modulates the behavior of GEV distribution.

We fit the nonstationary GEV distribution to the annual maximum river stage under the Bayesian framework. A full Bayesian MCMC provides a more complete representation of uncertainty of the parameter estimates with realistic credible intervals than is computationally feasible with the frequentist approach (Reis and Stedinger 2005). Bayes' theorem combines the prior knowledge regarding the model parameters with the information gained from the observational data (i.e., the likelihood function) into the posterior distribution of the model parameters $\theta$, given the data $(p(\theta|x))$:

$$p(\theta|\omega) \propto L(\omega|\theta)p(\theta), \qquad (3)$$

where $L(\omega|\theta)$ is the likelihood function and $p(\theta)$ is the prior distribution of random variable $\theta$. For a random variable $\omega = \{\omega_1, \omega_2, \ldots, \omega_N\}$, the likelihood function $L(\theta)$ for the parameter vector $\theta = (\mu_0, \mu_1)$ associated with its probability density function ($f$) is defined as:

$$L(\theta) = \prod_{i=1}^{N} f(\omega_i|\theta). \qquad (4)$$

where $i=1,2,\ldots,N$ indexes the years of river stage data.

We sample from the posterior distribution of the model parameters using the Metropolis-Hastings algorithm (Chib and Greenberg 1995). We use a Gaussian prior distribution with a large variance ($N(0,100)$) for each parameter on GEV model. We also explore the sensitivity of river stage to different prior distributions, including uniform prior, and Gaussian prior distribution with large variance ($N(0,100)$) and small variance ($N(0,1)$). We sample each GEV parameter successively for 100,000 iterations. The first 10,000 iterations are removed for burn-in. We use the remaining 90,000 samples to serve as the ensemble for analysis.

We incorporate potential nonstationarity into the GEV model by allowing the model parameter to covary linearly with Indian Ocean Dipole (IOD) index (see Supplementary material, Fig. A2). Future study could consider several other potential covariates to capture both local and large-scale dynamics, and further explore any nonlinear relationship between covariate and GEV parameters. Previous studies (Ajayamohan and Rao 2008, Ashok et al. 2001) have shown that different hydrometeorological events (e.g., extreme rainfall and floods) in south Asia are modulated by IOD. The IOD is quantified with Dipole Mode Index (Saji et al. 1999). The Dipole Mode Index is the anomalous sea-surface temperature gradient between the tropical western Indian Ocean (50 °E - 70 °E, 10 °S - 10 °N) and the tropical southeastern Indian Ocean (90 °E - 110 °E, 10 °S - Equator) (Saji et al. 1999). We use the historical monthly Dipole Mode Index derived from the HadISST dataset (https://psl.noaa.gov/gcos_wgsp/Timeseries/DMI/). We calculate the mean from June to November following Agilan and Umamahesh (2017) to use as input to the nonstationary model.

## 2.4. Sensitivity analysis

We perform sensitivity analysis of extreme river stage estimates on specification of different uncertainty sources, including model prior distributions, model structures and model parameters. Model prior uncertainty refers to the uncertainty contribution from the use of noninformative and informative sets of prior distributions for the model parameter. Model structural uncertainty refers to the uncertainty due to stationary and nonstationary GEV distributions. Parametric uncertainty represents the full ensemble in the parameter sample obtained using Bayesian MCMC approach. We note that these scenarios do not cover the set of all possible scenarios. Thus, the considered scenarios help to characterize key aspects of the uncertainties as opposed to fully quantify them.

We employ a cumulative uncertainty approach (Kim et al. 2019) to assess the sensitivity of extreme river stage estimates to different uncertainty sources. The cumulative uncertainty approach allows us to assess the individual as well as combined impacts of various uncertainty sources in extreme river stage estimates. We perform the sensitivity analysis from each source, called the individual uncertainty. Individual uncertainty is the sum of the variation of the main effect of source $z$ and the variations of the interactions between source $z$ and additional sources after $z$. To compute the individual source uncertainty, we first compute the conditional cumulative uncertainty up to a particular source. Conditional cumulative uncertainty up to a particular source is defined as the variation in the extreme river stage estimates due to the modeling choices up to that source, while the choices beyond that source are fixed. Then the marginal cumulative uncertainty up to a particular source is an average of conditional cumulative uncertainties. Finally, the individual uncertainty is the difference between successive marginal cumulative uncertainties.

Let, $Z$ be the uncertainty sources in extreme river stage estimate, where in our case $Z=3$, i.e., prior distributions, model structures and model parameters. For a particular uncertainty source $z$, there are $nz$ probabilistic scenarios denoted by $\chi_z$. The probabilistic scenarios are obtained by considering three sets of model priors, two sets of model structures, and ensemble of parameter samples from MCMC chain. The cumulative uncertainty due to different uncertainty sources is defined as the variation in the extreme river stage estimates due to the choice of scenarios up to source $z$, while the scenarios after source $z$ are fixed. The cumulative uncertainty up to source $z$ is denoted by $U^{cum}(\chi_1,...,\chi_z)$. For a specific scenario, of source $z$ for $z = 1,...,Z$, we let $P(x_1, x_2, ..., x_Z)$ be the extreme river stage estimates using the scenarios $\chi_z, z = 1,...,Z$. The set of extreme river stage estimates for a given scenario, after source $z$, are:

$$q_{x_{z+1},...,x_Z} = \{P(x_1,...,x_z,x_{z+1},...,x_Z): x_j \in \chi_j, j = 1,...,z\}. \quad (5)$$

Then $U^{cum}(q_{x_{z+1},...,x_Z})$ can be interpreted as the conditional cumulative uncertainty up to source $z$ while the scenarios after source z are fixed as $x_{z+1,...,}x_Z$. The marginal cumulative uncertainty up to source $z$ is the average of conditional cumulative uncertainties defined as:

$$U^{cum}(\chi_1,...,\chi_z) = \frac{1}{\prod_{j=z+1}^{Z} n_j} \sum_{x_{z+1} \in \chi_{z+1}} \cdots \sum_{x_Z \in \chi_Z} U(q_{x_{z+1},...,x_Z}). \quad (6)$$

Since the cumulative uncertainty is monotonously increasing (Kim et al., 2019), we can define the individual source uncertainty as the difference between successive cumulative uncertainties. That is, the uncertainty of source $z$, denoted by $U^{cum}(\chi_z)$, can be defined as:

$$U^{cum}(\chi_z) = U^{cum}(\chi_1,...,\chi_z) - U^{cum}(\chi_1,...,\chi_{z-1}) \quad (7)$$

Note that the uncertainty of each source is the amount of contribution to the cumulative uncertainty. Also, the sum of uncertainties of individual sources is always equal to the total uncertainty $U^{cum}(\chi_1,...,\chi_z)$. The uncertainty of stage $z$ can also be defined as the sum of the variation of the main effect of source $z$ and the variations of the interactions between source $z$ and additional uncertainty sources after $z$. The uncertainty of stage z can be expressed as

$$Variance\left(\beta_{\chi_z}^{(Z)}\right) + \sum_{h=z+1}^{Z} Variance\left(\beta_{\chi_z,\chi_h}^{(z,h)}\right), \quad (8)$$

where $\beta_{\chi_z}^{(Z)}, z = 1,...,Z$ is the main effect term; and $\beta_{\chi_z,\chi_h}^{(z,h)}$ represents the interaction effect.

We express both the individual and cumulative uncertainties in terms of the range (Chen et al., 2011) and variance (Bosshard et al., 2013) in extreme river stage estimates. Let, $Y = q_{x_{z+1},...,x_Z}$ and a set of $Y = \{y_1,...,y_n\}$. The range is defined as:

$$Range(Y) = max_{1 \leq i \leq n} y_i - min_{1 \leq i \leq n} y_i, \quad (9)$$

and the variance is defined as:

$$Variance(Y) = \frac{1}{n} \sum_{i=1}^{n} (y_i - \bar{y})^2, \quad (10)$$

where, $\bar{y} = \frac{1}{n}\sum_{i=1}^{n} y_i$.

# 3. Results

## 3.1. Change-point and trend analysis

We investigate the validity of the stationarity assumption in extreme river gauge records at the selected hydrologic stations (Fig. 2). The change point in the mean is significant at the 5% significance level in Benighat (Fig. 2d) and Rabuwabazar (Fig. 2f). These stations also depict significant change point in mean in annual maximum discharge records (see Supplementary material, Fig. S1). Note that both Rabuwabazar and Benighat represent headwater basins in the high-mountain region. The change-points can be the results of a single factor or combination of multiple factors such as changes in precipitation patterns, glacier melting, land-use/land-cover changes, water transfers, dam constructions and reservoir regulations (Alam et al., 2017; Gautam and Acharya, 2012; Nie et al., 2018; Talchabhadel et al., 2018). Engineering structures for flow regulation, such as dams and reservoirs, can contribute to the nonstationary character of the streamflow and might induce a shift in the magnitude of annual maximum floods. To evaluate the potential impacts of such structures on the streamflow regime, a dimensional index such as the reservoir index or check dam index can be used as an alternative covariate of the nonstationary model (Lopez and Frances 2013). Rabuwabazar (Fig. 2f) exhibit significant monotonic trends for the flood peak records prior to the change point. None of these two stations present a statistically significant trend after the change-point. For stations that lack significant change-point in mean, we perform temporal trend analysis on the entire time series. Stations including Arughat, Banga Belgaon, Asaraghat, Chisapani, Bagasoti and Narayanghat do not exhibit a statistically significant change-point in mean. In addition, these stations do not show statistically significant temporal trend on the entire flood peak records (Fig. 2). As suggested in other studies (Villarini et al., 2011; Villarini and Smith, 2010), we notice that change-points rather than monotonic trends are responsible for nonstationarity in extreme river gauge records.

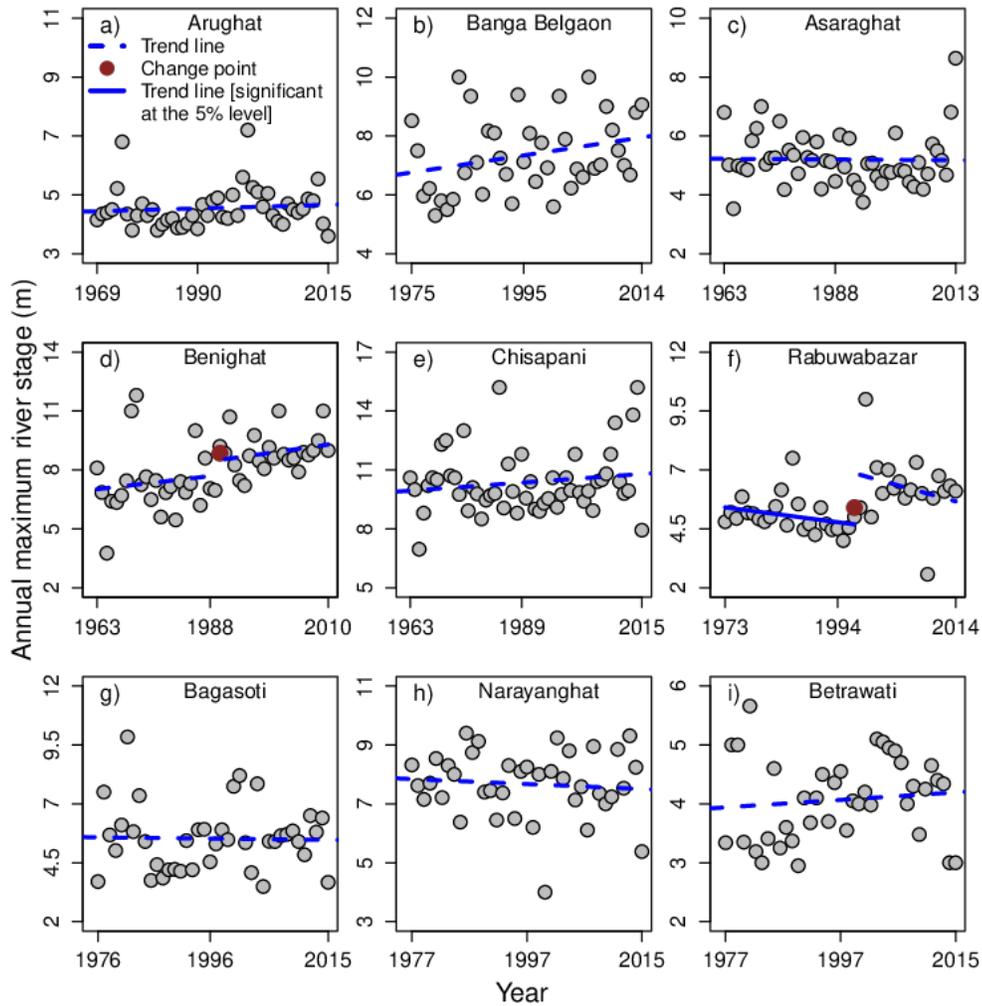

**Figure 2**. Instantaneous annual river gauge records (gray circle) in the selected hydrologic stations. Brown circle shows the change point. We show the trend line in the overall record, as well as in the record before and after the change point. Solid lines represent that the trend is significant and dashed lines represent that the trend is insignificant at the 5 % significance level.

## 3.2. River stage estimates

We use stationary model structure to estimate river stage in hydrologic stations (Arughat, Asaraghat, Chisapani, Banga Belgaon, Bagasoti and Narayanghat) that do not exhibit statistically significant change-point in mean and temporal trends (Fig. 3 and refer Supplementary material, Fig. S3). We use MCMC sampling within a Bayesian framework to estimate the GEV parameter. We adopt the MCMC sample with the highest posterior probability samples as the "best guess" estimate of that parameter, referred to as maximum posteriori estimates of the model parameter. We consider the full ensemble of samples to account for the uncertainty of flooding frequency.

We find that neglecting model parametric uncertainty underestimates the expected river stage (Fig. 3). Using the maximum a posteriori estimates of model parameters, as opposed to the full parameter sample, underestimates the extreme river stage by as much as 7% at 100-year return period (Fig. 3). This underestimation bias is driven by the right-skewed return level distribution where the mode (6.9 m) is smaller than the mean (7.4 m) (Fig. 3b). In addition, compared to the expected return river stage from the Bayesian MCMC approach, the commonly used frequentist approach underestimates the river stage. This underestimation can drive higher flood hazards.

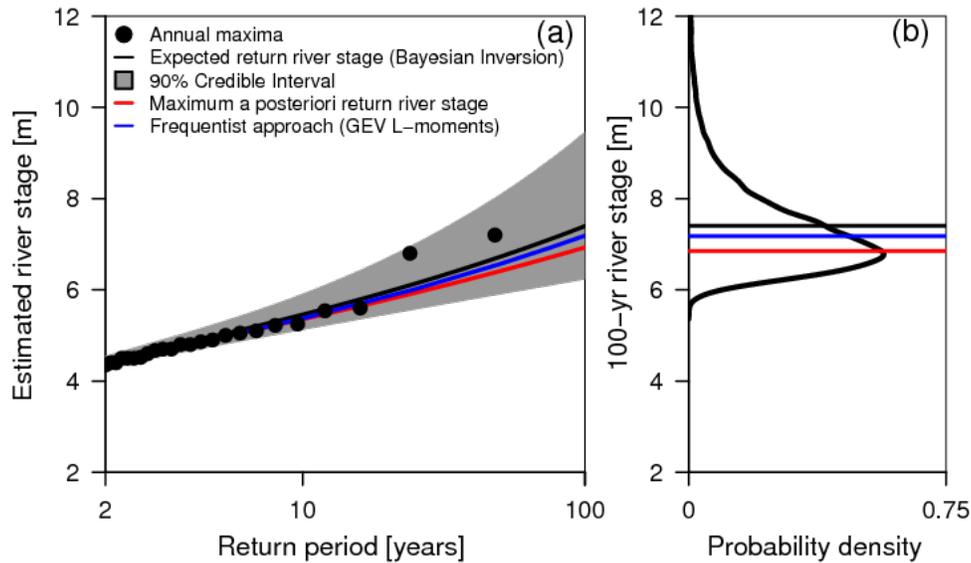

**Figure 3**. Estimated river stage in Arughat station: (a) Estimated river stage considering uncertainty (black line and gray bounds) and ignoring uncertainty (red line) using stationary model structure, and (b) comparison of different estimates of the 100-yr river stage. Note that a 100-year river stage is a flood event that has a 1 in 100 chance (1% probability) of being equaled or exceeded in any given year. We use a Gaussian prior distribution with a wide variance ($N(0,100)$) for each parameter on the GEV model.

With the observed changes in river gauge records, extreme value analysis at Benighat and Rabuwabazar require considering nonstationarity in the probability distribution of river stage records (Fig. 2 and Fig. 4). Given the nonstationarity in river gauge records, the stationary assumption leads to underestimation of extreme river stage (Fig. 4). The downward bias under stationary assumption increases with higher return periods. Also, the uncertainty increases as the return period increases. This feature is evident in both hydrologic stations Rabuwabazar (Fig. 4) and Benighat (see Supplementary material, Fig. S4). For example, for a flood event with a return period of 25-years at Rabuwabazar (Fig. 4a), the difference between the nonstationary and stationary extreme flood estimates is about 0.75 m (+9%); while for a 100-year return period event, the difference between nonstationary and stationary river stage estimates is over 1.5 m (+ 16%). At Rabuwabazar station, which represents a relatively small basin size (3650 km$^2$) and steep slope (basin average slope is 22%), this additional 1.5 m increase in river stage can be the difference of millions of dollars in potential damages across the sectors. This risk is likely to increase in the future under changing climatic conditions. There is strong evidence of climate change to alter regional hydrologic processes, for instance, atmospheric warming and altered monsoon rainfall patterns is likely to result in higher snowmelt and extreme flood events in the future (Karki et al. 2020; Devkota and Gyawai, 2015; Poudel et al. 2020).

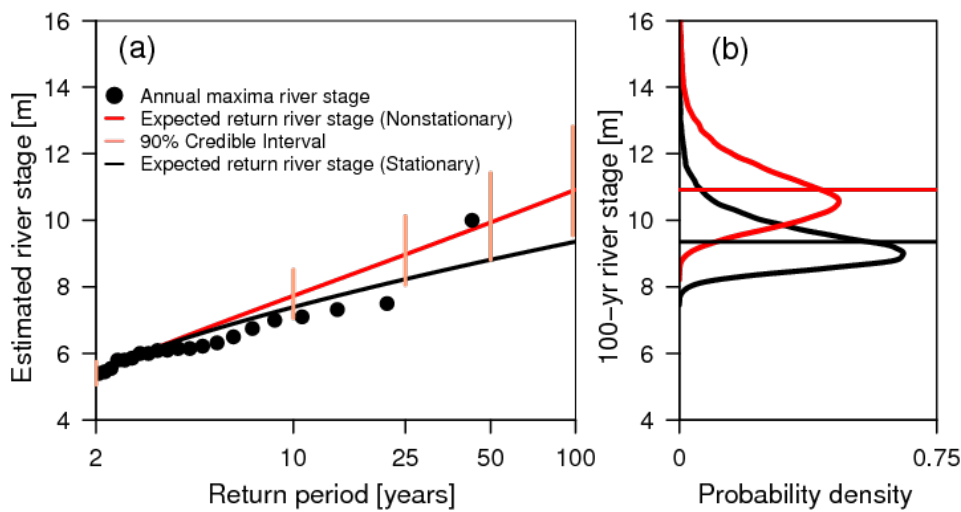

**Figure 4**. Estimated river stage in Rabuwabazar station: (a) Estimated river stage under stationary and nonstationary model structures, and (b) comparison of different estimates of the 100-yr river stage.

If a stationary river stage estimate is used to inform critical infrastructure design, the infrastructure cannot withstand the extreme events in changing climatic conditions. This is shown by the nonstationary extreme river stage estimates for the same return period. For historical river gauge records with a stationary return period of 100-years, the corresponding return periods under nonstationary conditions at Rabuwabazar (Fig. 4) and Benighat (see Supplementary material, Fig. S4) reduce to 33, and 15 years, respectively. The reason for the bigger differences in the return period in Benighat (see Supplementary material, Fig. S4) could be because of the sharp increasing trend of river stage compared to other stations (Fig. 2d). As a consequence, to ensure a 15-year level of protection under nonstationary conditions, one may have to use an initial design that has a higher level of protection under stationary assumption.

We further demonstrate the impact of varying prior distributions of the parameters in river stage estimates (Fig. 5). We find that the estimated river stage relies on the contribution from prior distribution of parameters, particularly in the upper tail of the probability distribution. This is illustrated by the survival functions (Fig. 5b). For a given prior, the 100-year return level estimate for a nonstationary model is larger than the stationary model, and further exhibit wider credible interval compared to the stationary estimates. For a given model structure, the changes in river stage using different priors is noticeable in the tails of the distribution (Fig. 5b). At the 1/100 level (Fig. 5b), the choice of priors can lead to the differences in nonstationary river stage estimates as high as 0.6m. In addition, the resulting posterior estimates from gaussian priors exhibit tighter credible intervals relative to the uniform priors. For example, the 90% credible interval in nonstationary extreme river stage estimates using uniform prior is 9.54 - 13.04 m, and the resulting expected return river stage is 10.98 m. However, the gaussian priors with small variance expectedly tightens these estimates slightly, with a 90% credible interval of 9.35 - 12.57 m and expected return river stage of 10.69 m.

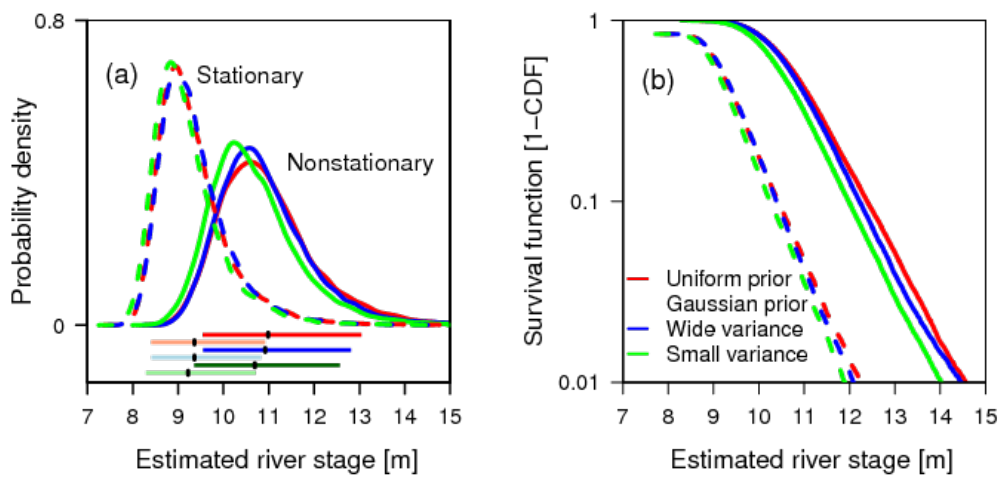

**Figure 5**. Estimated 100-year river stage in Rabuwabazar: (a) Posterior probability density function (pdf) of the 100-year river stage using two different model structures (stationary in dashed lines and nonstationary in solid lines) and three different prior distributions, and (b) survival function. Below the pdfs are boxplots

for different model structures and prior distributions. The bold vertical lines denote the ensemble medians; the horizontal bars denote the 5% - 95% quantile range.

### 3.3. Sensitivity analysis

Sensitivity analysis in river stage return levels is critical to better communicate flood risk, identify ways to improve the reliability of extreme value estimation and guide future research efforts for uncertainty reduction. We quantify the sensitivity of river stage return level estimates to key uncertainty sources, including model structures, prior distributions and model parameters (Fig. 6). We consider three sets of model priors, two sets of model structures, and ensemble of parameter samples from MCMC chain. In Fig. 6, we show the sensitivity analysis results for Rabuwabazar station.

Total uncertainty in the extreme river stage estimates is the contribution from each individual source and their interactions. We find that the model parameter dominates the decomposition of range and variance in extreme river stage estimates. Model parameters contribute more than the combined effect of model structures and prior distributions (Fig. 6). The longer the return period, the larger the contribution from the model parameter. Therefore, to a specific model structure, it is critical to use the most suitable method to estimate the parameters of the probability distribution. If the GEV distribution with nonstationary location, scale and shape parameters is to be considered, limited data records are not sufficient for a good inference, especially in the shape parameter, and hence in the extreme river stage estimates (Ragulina and Reitan 2017). Other statistical approaches (e.g., peaks-over-thresholds with Poisson process/generalized Pareto distribution (Wong et al. 2018); metastatistical extreme value distribution (Miniussi et al. 2020)) or process-based approaches (e.g., Perez et al. 2019) that make use of more data than the annual maximum river stage records may be of use to constraint the shape parameter.

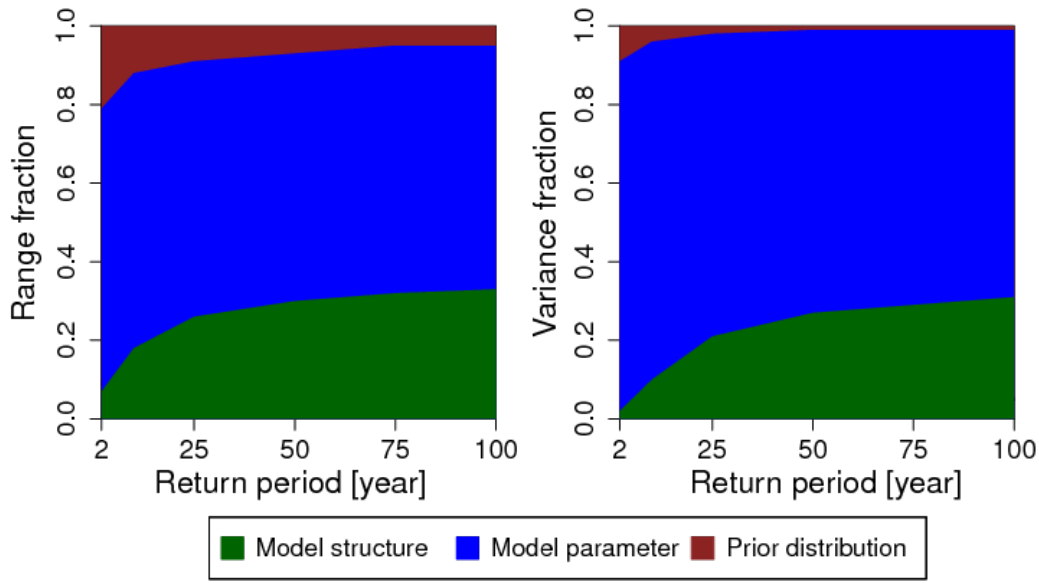

**Figure 6**. Range and variance decomposition in return river stage estimates in Rabuwabazar station. We perform sensitivity analysis on specifications of model structure, prior distribution, and model parameter. We consider stationary and nonstationary GEV model structures. We consider parameter samples from MCMC chain. We analyze both noninformative uniform prior distribution, as well as informative sets of prior distributions such as Gaussian prior distribution with a wide variance ($N(0,100)$) and narrow variance ($N(0,10)$).

After the model parameter, model structure plays the most important role. The impact of model structure is evident at longer return periods. For example, the range and variance explained by the model structure dampens and contributes less than 10% of the total uncertainty at 2-year return period, whereas more than 30% of total sensitivity is contributed by model structure at 100-year return period. This suggests that the sensitivity analysis should be conducted for return periods of interest independently from other return periods (Ward et al. 2011). Critical infrastructures such as dams and reservoirs generally have longer service life, and hence require consideration of design flood with longer return period to withstand the extreme events over the lifetime. Thus, the appropriate choice of model structure is important for resilient infrastructure design in a changing climate.

We find that the selected prior distributions exhibit relatively small contributions compared to any other considered uncertainty source. Note that these contributions only represent relative influence but not absolute impacts. In the case of longer return periods, the contribution of model priors is different for the two uncertainty measures. For example, when we use the range as the sensitivity measure, about 10% of total sensitivity is contributed by model priors at 100-year return period; whereas the variance explained by the model priors dampens and contributes less than 5% of the total uncertainty. This highlights the importance of using different measures for uncertainty decomposition. The decision to choose the prior distribution for the parameter depends on any available knowledge about the parameters. Previous studies (Lee et al. 2017) have shown that if sufficient data record is available, the parameter estimates are not very sensitive to the choice of prior distributions. In such cases, the information contained in the dataset is dominant. However, in the case of limited flood peak records, the choice of the reasonable prior distribution such as informative prior distribution can become important.

## 4. Discussion and Conclusion

We present here a Bayesian-based framework to incorporate nonstationarity into extreme river stage estimates and use this framework to quantify the sensitivity of river stage estimates to key uncertainty sources. We consider three different uncertainty sources: model structures, model parameters and prior distribution of the parameters. We implement the framework in nine major river basins in the central Himalayan region, Nepal. We find that the choice of methods and assumptions used in the data-driven statistical models can modulate the estimates. We show that neglecting key uncertainties surrounding extreme value analysis can mislead to hazard estimates. Given nonstationarity in gauge records, stationary assumption can substantially underestimate river stage return levels particularly at longer return periods. If such an extreme flood under stationary assumption is used to inform design strategies, the infrastructure cannot withstand the extreme flood events over the anticipated lifetime. We find that the considered model parametric uncertainty is more influential than model structures and prior distributions in extreme river stage estimates. Overall, our results highlight three key points: i) current approaches to estimate flood hazards neglect key known unknowns such as model structural and parametric uncertainties; ii) neglecting uncertainties can drastically underestimate extreme flood hazard probability distribution; and iii) comprehensive uncertainty characterization is critical to guide the future efforts of uncertainty reduction and communicate decision-relevant extreme information.

Reliable estimates of flood frequency and magnitude are essential for infrastructure planning, design and minimization of associated risks in a changing climate. These estimates are generally provided as regional and/or national standards in the form of design manuals or codes (Atlas 2020). Such standards provide guidance to engineers to size infrastructures so that acceptable performance could be achieved over the service life (Brown et al. 2009). Design guidelines could ensure uniformity in extreme estimates to be used in local-to-national level planning decisions. Apart from the design of infrastructures, such specifications could be used to inform flood management and mitigation strategies, depict floodplain boundaries that are most vulnerable to floods, and manage development activities in the floodplain. However, such guidelines are currently lacking in Nepal. Thus, comprehensive studies on hydroclimatic extremes considering a wider range of physical drivers, hydroclimatic regimes, basin sizes, catchment characteristics and longest possible records are required for establishing reliable engineering standards and design codes.

Hydropower provides a huge contribution to Nepal's economy. Nepal has the economically viable potential to put in place over 40,000 megawatts of hydropower generation capacity (Alam et al. 2017). However, the current installed hydro capacity is less than 1,000 megawatts (Hussain et al. 2019). Nepal is going through a rapid hydro-infrastructure development to meet its energy demand (Fig. 1). As such, nonstationary extreme river stage estimates are critical to inform hydro-infrastructure decision-making under climate change. However, designing infrastructure or maintaining reliable levels of hydropower delivery in the face of deep and dynamic climate uncertainties pose highly complex decision problems. Different approaches such as robust infrastructure decision making under deep uncertainty (Herman et al. 2014) can provide insights into potential robust strategies, characterize the vulnerabilities of such strategies, and evaluate trade-offs among them.

Considering that the design life of much of the flood-sensitive infrastructures (e.g., bridge piers and emergency spillway) is greater than 100-years, careful attention should be given on selecting appropriate design flood in changing climatic conditions. Apparently, larger design return periods lead to robust infrastructures with smaller failure risk, but such infrastructures would also require higher upfront investment. It is ultimately up to the decision makers to quantify the tradeoffs between the upfront investment, level of protection it provides over the service life and consequences of failure. Thus, critical infrastructure design in a changing climate requires an integrated approach, as the decisions are informed by several disciplines, including hydrology, civil engineering, climate science, economics and decision-science.

Extreme floods in the Himalayas are caused by a mixture of flood-generating mechanisms, with prolonged summer monsoon rainfall, localized cloud bursts, sudden glacial lake outburst and/or landslide lake outburst playing a central role (Huss et al. 2017, Schwanghart et al. 2016, Ragettli et al. 2016, Veh et al. 2020). Since each individual event is unique, more studies involving physical mechanisms are required to fully characterize flood hazards and infrastructure vulnerability to climate change. Future research could be focused on exploring changes in climate variables and evidence of human activity that contributes to abrupt changes, trends, and nonstationarity in extreme water level. In addition, flood risk management requires nonstationary flood risk projections from future climatic conditions. Incorporating these changing conditions into design specifications have implications to improve the reliability of infrastructure over the service life.

Under review in Hydrological Sciences Journal